# Engineering Plateau Phase Transition in Quantum Anomalous Hall Multilayers

Deyi Zhuo[1,3], Ling-Jie Zhou[1,3], Yi-Fan Zhao[1], Ruoxi Zhang[1], Zi-Jie Yan[1], Annie G. Wang[1], Moses H. W. Chan[1], Chao-Xing Liu[1], Chui-Zhen Chen[2], and Cui-Zu Chang[1]

[1]Department of Physics, The Pennsylvania State University, University Park, PA 16802, USA

[2]Institute for Advanced Study and School of Physical Science and Technology, Soochow University, Suzhou 215006, China

[3] These authors contributed equally: Deyi Zhuo and Ling-Jie Zhou

Corresponding authors: cxc955@psu.edu (C.-Z. Cha.); czchen@suda.edu.cn (C.-Z. Che.).

**Abstract: The plateau phase transition in quantum anomalous Hall (QAH) insulators corresponds to a quantum state wherein a single magnetic domain gives way to multiple domains and then re-converges back to a single magnetic domain. The layer structure of the sample provides an external knob for adjusting the Chern number *C* of the QAH insulators. Here, we employ molecular beam epitaxy (MBE) to grow magnetic topological insulator (TI) multilayers with an asymmetric layer structure and realize the magnetic field-driven plateau phase transition between two QAH states with odd Chern number change Δ*C*. In multilayer structures with *C*=±1 and *C*=±2 QAH states, we find characteristic power-law behaviors between temperature and the scaling variables on the magnetic field at transition points. The critical exponents extracted for the plateau phase transitions with Δ*C*=1 and Δ*C*=3 in QAH insulators are found to be nearly identical, specifically, $\kappa_1$~0.390±0.021 and $\kappa_2$~0.388±0.015, respectively. We construct a four-layer Chalker-Coddington network model to understand the consistent critical exponents for the plateau phase transitions with Δ*C*=1 and Δ*C*=3. This**

**work will motivate further investigations into the critical behaviors of plateau phase transitions with different Δ*C* in QAH insulators and provide new opportunities for the development of QAH chiral edge current-based electronic and spintronic devices.**

**Main text:** The tuning of the quantum phase transition between two different topological states is a subject of enduring interest [1,2]. A classic example is the plateau phase transition in the quantum Hall (QH) effect [3-6]. The QH liquid and Anderson/normal insulator phases associated with such a phase transition are usually examined by analyzing the temperature-dependent transport behaviors and found to be characterized by a single critical exponent $\kappa$ [4-6]. The quantum anomalous Hall (QAH) effect is a zero magnetic field manifestation of the QH effect [2]. The QAH effect, usually realized by time-reversal symmetry breaking in topological nontrivial systems [2,7-11], harbors spin-polarized dissipationless chiral edge states with vanishing longitudinal resistance and quantized Hall resistance of $h/Ce^2$. Here $C$ is the Chern number that corresponds to the number of chiral edge channels [12,13]. The QAH effect was envisioned by Duncan Haldane [7] and experimentally realized in molecular beam epitaxy (MBE)-grown magnetically doped topological insulator (TI) thin films, specifically Cr- and/or V-doped $(Bi,Sb)_2Te_3$ [2,9,13-19]. To date, considerable experimental efforts have explored the global phase diagrams and scaling behaviors of the plateau phase transition in QAH insulators with Chern number change Δ*C*=2 [15,18,20-24]. The exponent $\kappa$ found in these studies with different samples varies over a broad range between 0.22 and 0.62 [2,20-24]. The transition with a larger Δ*C* in QAH and in QH states have seldom been studied. Therefore, the creation of plateau phase transitions with different Δ*C* in a single magnetic TI multilayer sample provides a new platform for the exploration of the critical behaviors of plateau phase transitions.

Recently, the QAH effect with Chern number *C* from 1 to 5 has been realized in heavily Cr-doped TI/TI multilayers [13,25]. In these multilayers, the heavily Cr-doped TI layer acts as a trivial ferromagnetic insulator, gapping out the surface states of TI layers and effectively binding several $C=1$ QAH layers together. Moreover, since both Cr- and V-doped $(Bi,Sb)_2Te_3$ films exhibit the QAH effect [9,14], it is feasible to create plateau phase transitions with different $\Delta C$ by inserting (Cr,V)-co-doped $(Bi,Sb)_2Te_3$ layers with different coercive field $\mu_0H_c$ into [Cr-doped $(Bi,Sb)_2Te_3/(Bi,Sb)_2Te_3]_m$ multilayers (Figs. 1a and S9a) [26].

In this work, we employ MBE to fabricate magnetic TI multilayer with an asymmetric structure, specifically 3 quintuple layers (QL) V-doped $(Bi,Sb)_2Te_3$/[3QL $(Bi,Sb)_2Te_3$/2QL Cr-doped $(Bi,Sb)_2Te_3]_m$ (Fig. 1a). Here $m$ is the number of the 3QL $(Bi,Sb)_2Te_3$/2QL Cr-doped $(Bi,Sb)_2Te_3$ bilayer. We perform electrical transport measurements on the $m=2$ sample and realize magnetic field-induced plateau phase transitions with $\Delta C=1$ and $\Delta C=3$ in QAH insulators (Figs. 1b and 1c). Remarkably, despite the distinct $\Delta C$, the power-law scaling analysis yielded identical critical exponent $\kappa$. We provide a 2D Dirac effective Hamiltonian and a four-layer Chalker-Coddington network model [27-29] to describe the QAH plateau phase transitions with odd $\Delta C$. The consistent critical exponents in both theoretical calculations and experimental results are expected to be identical to that of QH systems during the QAH plateau phase transitions with odd $\Delta C$.

We first perform electrical transport measurements on the $m=2$ multilayer at $V_g=V_g^0=+15V$ and $T=25mK$ (Fig. 1c). The value of the charge neutral point $V_g^0$ here is determined when the zero magnetic field longitudinal resistance $\rho_{xx}$ is minimized. Note that the 3QL $(Bi,Sb)_2Te_3$ layer between two heavily Cr-doped $(Bi,Sb)_2Te_3$ layers is thick enough to reduce the interlayer coupling [30]. When these two layers have a parallel magnetization alignment, the sample shows a $C=2$ QAH state with a pair of chiral edge channels (Fig. 1b). The value of $\rho_{yx}$ is found to be $\sim 0.505h/e^2$ at $\mu_0H=0T$, concomitantly with $\rho_{xx} \sim 0.007h/e^2$ (Figs. 1c and 1d).

Besides the $C=2$ QAH effect, the $m=2$ multilayer also shows the $C=-1$ QAH state with a single chiral edge channel when the top V-doped $(Bi,Sb)_2Te_3$ and bottom two Cr-doped $(Bi,Sb)_2Te_3$ layers have an antiparallel magnetization alignment (Fig. 1b). The values of $|\rho_{yx}|$ and $\rho_{xx}$ are found to be $\sim 1.005h/e^2$ and $\sim 0.021h/e^2$ at $\mu_0H=0.5$T, respectively (Figs. 1c and 1d). In this scenario, the $C=-1$ QAH effect can be considered as arising from the simultaneous existence of an axion insulator state and a parallel $C=-1$ QAH state. We note that both $C=1$ and $C=2$ QAH states can persist at $\mu_0H=0$T through minor loop measurements (Fig. S3) [26]. Moreover, both the $C=2$ and $C=1$ QAH states in the $m=2$ multilayer are further demonstrated by the $(V_g-V_g^0)$ dependence of $\rho_{yx}$ and $\rho_{xx}$ at $\mu_0H=0$T and $\mu_0H=0.5$T after magnetic training [labeled as $\rho_{yx}(0)$ and $\rho_{yx}(0.5T)$, Fig. 1d]. We find that $\rho_{yx}(0)$ and $\rho_{yx}(0.5T)$ exhibit two quantized plateaus near $V_g=V_g^0=+15$ V with the quantized values of $\sim h/2e^2$ and $\sim h/e^2$, respectively. Concomitantly, $\rho_{xx}(0)$ and $\rho_{xx}(0.5T)$ show zero resistance plateaus, confirming the well-quantized $C=2$ and $C=1$ QAH states in the $m=2$ multilayer.

When $\rho_{yx}$ and $\rho_{xx}$ of the $m=2$ multilayer are converted into conductance. $\sigma_{xx}$ shows four peaks near $\pm\mu_0H_1$ ($\sim\pm0.906$T) and $\pm\mu_0H_2$ ($\sim\pm0.204$T), corresponding to the magnetization reversal in the top V-doped $(Bi,Sb)_2Te_3$ and bottom two Cr-doped $(Bi,Sb)_2Te_3$ layers, respectively (Fig. 2a). In the well-defined magnetization regimes, $|\sigma_{xy}|$ exhibits quantized values of $\sim e^2/h$ and $\sim 2e^2/h$, signifying the QAH states with $C=1$ and $C=2$ (Fig. 2b). The plateau phase transitions between QAH states with different $C$ can be examined by plotting the flow diagram $(\sigma_{xy}, \sigma_{xx})$ with sweeping $\mu_0H$ (Fig. 2c). During the plateau phase transition, the behavior of $(\sigma_{xy}, \sigma_{xx})$ traces a semicircle curve, locating quantum critical regimes [15,18,20,24,31]. The $C=\pm 1$ and $C=\pm 2$ QAH states correspond to $(\sigma_{xy}, \sigma_{xx}) = (\pm e^2/h, 0)$ and $(\pm 2e^2/h, 0)$, respectively (Fig. 2c). We note that the radius of the semicircle curve is determined by $\Delta C$ of the plateau phase transition in QAH insulators. In the $m=2$ multilayer, the radii of the two semicircles are $e^2/2h$ and $3e^2/2h$, confirming two kinds of plateau phase transitions with $\Delta C=1$ and $\Delta C=3$.

Besides sweeping $\mu_0H$, we also plot the flow diagram ($\sigma_{xy}$, $\sigma_{xx}$) of the $m=2$ multilayer by sweeping $V_g$ under different $\mu_0H$ (Fig. 2d). For $\mu_0H=\pm1.5$T and $\mu_0H=\pm0.5$T, the evolution begins with a large $\sigma_{xx}$ at $V_g=100$V towards the corresponding $C=\pm2$ and $C=\pm1$ QAH states at $V_g=V_g^0$. As $V_g$ decreases from $V_g^0$ to -60V, all [$\sigma_{xy}(V_g)$, $\sigma_{xx}(V_g)$] trajectories under different $\mu_0H$ tend to deviate to the fixed QAH points at ($\pm 2e^2/h$, 0) and ($\pm e^2/h$, 0) covering the continuous semicircle curves with a radius of $3e^2/2h$ centered at ($\sigma_{xy}$, $\sigma_{xx}$) = ($\pm e^2/2h$, 0) (Fig. 2d). The universal scaling behaviors here resemble the quantum criticality of the plateau phase transitions between adjacent QH states and QAH states [15,32].

Next, we focus on the scaling behaviors of the two plateau phase transitions with $\Delta C=1$ and $\Delta C=3$. At $V_g=V_g^0$, we perform $\mu_0H$ dependence of $\rho_{xx}$ and $\rho_{yx}$ under 50mK$\leq T \leq$300mK (Figs. 3a and 3b). Within this temperature range, the $C=2$ QAH state near $\mu_0H=0$T remains relatively stable. The value of $\rho_{yx}(0)$ changes from ~0.492$h/e^2$ at $T=50$mK to ~0.486$h/e^2$ at $T=300$mK, and the value of $\rho_{xx}(0)$ correspondingly increases from ~0.0008$h/e^2$ at $T=50$mK to ~0.032$h/e^2$ at $T=300$mK. However, for the $C=1$ QAH state near $\mu_0H=0.4$T, the value of $\rho_{yx}(0.4\text{T})$ deviates from ~0.967$h/e^2$ at $T=50$mK to ~0.924$h/e^2$ at $T=300$mK, and the value of $\rho_{xx}(0.4\text{T})$ correspondingly increases from ~0.008$h/e^2$ at $T=50$mK to ~0.154$h/e^2$ at $T=300$mK. We note that while both the $C=1$ QAH state under antiparallel magnetization alignment and the $C=2$ QAH state under parallel magnetization alignment have chiral edge transport, their distinct sensitivities to temperature, as reflected in their $\rho_{yx}$ and $\rho_{xx}$ responses, might be due to the coexistence of the axion insulator state and the $C=1$ QAH state under antiparallel magnetization alignment.

To investigate the scaling behaviors of the two plateau phase transitions with $\Delta C=1$ (from $C=2$ to $C=1$ QAH states) and with $\Delta C=3$ (from $C=-2$ to $C=1$ QAH states), we show the $\mu_0 H$ dependence of $\rho_{xx}$ near the critical magnetic field $\mu_0 H_{c1}$ (Fig. 3c) and of $\rho_{yx}$ near the critical magnetic field $\mu_0 H_{c2}$ at 50mK$\leq T\leq$300mK (Fig. 3d), respectively. We find that all the $\rho_{xx}$ curves cross each other at a single point $\mu_0 H_{c1}\sim$0.97T (Fig. 3c), while all the $\rho_{yx}$ curves cross at $\mu_0 H_{c2}\sim$0.20T (Fig. 3d). These observations point to the presence of two quantum critical points between the $C=\pm 2$ and $C=1$ QAH states. The insulating ground state further confirms the coexistence of the axion insulator state and the $C=1$ QAH state. We note that for the plateau phase transition with $\Delta C=3$ near $\mu_0 H_{c2}$, the $\mu_0 H$ dependence of $\rho_{xx}$ is not employed in our scaling analysis. This is primarily due to its extremely sharp transition, which precludes a proper scaling analysis [1,2,24].

We first perform the scaling analysis of the plateau phase transition with $\Delta C=1$ from $C=2$ to $C=1$ QAH states (Fig. 3e). We find that the temperature dependence of the scaling variable $(\frac{d\rho_{xx}}{d\mu_0 H})_{\mu_0 H=\mu_0 H_{c1}}$ on the log scales shows a linear behavior for 100mK$\leq T\leq$300mK with a critical exponent $\kappa_1\sim$0.390$\pm$0.021 (Fig. 3e). For plateau phase transition with $\Delta C=3$ from $C=-2$ to $C=1$ QAH states, we find a similar critical exponent $\kappa_2\sim$0.388$\pm$0.015 (Fig. 3f). The values of both $\kappa_1$ and $\kappa_2$ are consistent with the prior value $\kappa\sim$0.38 found for the plateau phase transition between QAH and axion insulator states with $\Delta C=1$ [24]. Since power-law critical behaviors of these two QAH plateau phase transitions came from the same sample, it excludes the possibility of any effect due to different disorders of the interference between magnetic and electronic phase transitions that may be present in prior experiments. We note that both $(\frac{d\rho_{xx}}{d\mu_0 H})_{\mu_0 H=\mu_0 H_{c1}}$ and $(\frac{d\rho_{yx}}{d\mu_0 H})_{\mu_0 H=\mu_0 H_{c2}}$ start deviating from the linear dependence at and near $T=$100mK (Figs. 3e and 3f). It is likely that below 100 mK the electron temperature of the $m=2$ multilayer begins to deviate from that given by the thermometer located in the mixing chamber (Fig. S6) [26].

Next, we present a theoretical framework to understand the origins of the two plateau phase transitions. In the $m=2$ multilayer, there are four interfaces, and the effective Hamiltonian describing the properties of the $n^{\text{th}}$ interface is expressed as follows:

$$H_{\text{n}} = \hbar v_{\text{F}}(s_x k_y - s_y k_x)(-1)^n + M_n s_z \qquad (1)$$

The symbols $s_{x,y,z}$ represent Pauli matrices that operate within the spin space, while $k_{x,y}$ denote wave vectors in the $x$ and $y$ directions, respectively. $v_{\text{F}}$ stands for the Fermi velocity. In addition, the Dirac mass term $M_n$ represents the exchange field in the $z$-direction resulting from magnetization, with $n$ ranging from 1 to 4. $\sigma_{xy}^n = \text{sign}(M_n)e^2/2h$ for each surface is determined by the polarity of the magnetization $\text{sign}(M_n)$. Therefore, for the plateau phase transition with $\Delta C=1$ near $\mu_0 H_{c1}$, the magnetization reversal occurs merely on one interface between the top 3QL V-doped $(Bi,Sb)_2Te_3$ and 3QL $(Bi,Sb)_2Te_3$ layers. However, for the plateau phase transition with $\Delta C=3$ near $\mu_0 H_{c2}$, the magnetization reversal occurs across the three interfaces between 3QL $(Bi,Sb)_2Te_3$ and bottom 2QL Cr-doped $(Bi,Sb)_2Te_3$ layers (Figs. 1a and 1b). During this magnetization reversal, the Dirac mass term $M_n$ becomes spatially inhomogeneous, resulting in a 2D random Dirac mass Hamiltonian. This allows mapping onto the Chalker-Coddington model, elucidating the QH plateau phase transition [27-29]. Therefore, despite the difference in the magnetization reversal process, the values of the critical exponent $\kappa$ are similar for these two plateau phase transitions with $\Delta C=1$ and $\Delta C=3$.

To quantitatively illustrate the phase transitions, a four-layer Chalker-Coddington network model [33] is depicted on a square lattice structure (Fig. 4a) with intra-layer scattering $T_n$ and interlayer scattering $U$ [26]. In Fig. 4b, renormalized localization length $\Lambda$ is plotted against the scattering $\theta$, showing critical points $\theta_{c1}=0.46$ and $\theta_{c2}=1.38$ for the plateau phase transitions with $\Delta C$=1 and $\Delta C$=3, respectively. All data points near the critical point collapse onto a single curve (Figs. 4c and 4d inset) using a scaling function f($L/\xi$) with the correlation length $\xi$. Furthermore, fitting ln $\xi$ to the linear function of ln $|\theta - \theta_c|$ unveils slopes of $v_1$=2.69±0.31 and $v_2$=2.78±0.20 (Figs. 4c and 4d). These results support the finding that two plateau phase transitions with $\Delta C$=1 and $\Delta C$=3 share the same universality class as the plateau phase transition between QAH and axion insulator states [24] or the QH plateau phase transition [34-36]. The critical exponent $\kappa$=$p/2v$ convolutions of correlation length exponent $v$ and temperature exponent $p$ [37,38]. Consequently, for the two plateau phase transitions with $\Delta C$=1 and $\Delta C$=3 in the $m$=2 multilayer, the values of $p$=$2\kappa v$ are ~2.10±0.35 and ~2.16±0.24, respectively. The exponent $p$~2 surpassing the noninteracting value of $p$=1 might be influenced by the long-range Coulomb interactions [39], because of the low density of states of Dirac fermions. As the value of $p$ implies the inelastic scattering length, the nearly uniform values of $p$ across different quantum phase transitions lend support to the notion that a singular Hamiltonian can describe the different plateau phase transitions in a QAH insulator.

As noted above, prior studies on QAH plateau phase transitions [2,20-24] have indicated values of $\kappa$ ranging from 0.22 to 0.62 for the QAH plateau phase transition with $\Delta C$=2. This variability may be due to the sample-dependent disorder that cannot be avoided [2]. This work confirms that the two plateau phase transitions with different $\Delta C$ in the same QAH sample can be described by the Chalker-Coddington network models and an identical $\kappa$ is found. We note that plateau phase transitions with $\Delta C$=1 and $\Delta C$=3 in the $m$=2 multilayer fall within the same universal class as those in magnetic TI sandwiches [24], but differ from those in individual magnetic TI films [2,21-23]. In individual magnetic TI films, the coupling among distinct Dirac surface/interface states is more pronounced and collectively exerts a stronger influence. The plateau phase transition with $\Delta C$=2 in individual QAH films is characterized by a two-channel Chalker–Coddington model [40,41], suggesting a Berezinskii–Kosterlitz–Thouless-type transition [41]. However, in QAH multilayers, the presence of weak interlayer coupling results in plateau phase transitions with $\Delta C$=1 and $\Delta C$=3 sharing the same universality class as the one-channel Chalker–Coddington model [24,27]. So far, the impact of interlayer coupling on plateau phase transitions remains an open question, while a detailed exploration of this aspect is beyond the scope of our current study.

To summarize, we use MBE to fabricate magnetic TI multilayers with an asymmetric structure and realize the magnetic field-induced plateau phase transition with $\Delta C$=1 and $\Delta C$=3 in QAH insulators. By analyzing the scaling behaviors of these two plateau phase transitions, we find that these two plateau phase transitions share the same power law exponent. Our observation extends the larger global phase diagram and scaling behaviors of the QAH effect, which would predict that the QAH state would be connected continuously to other integer and fractional QAH states [42-45] and the corresponding phase transitions would belong to the same universality class in the Anderson transition regime. Our work paves the way for further explorations of the critical behaviors of plateau phase transitions between two QAH states with different Chern numbers.

**Acknowledgments:** This work is primarily supported by the ARO Award (W911NF2210159), including MBE growth and dilution transport measurements. The PPMS measurements are supported by the NSF grant (DMR-2241327) and the AFOSR grant (FA9550-21-1-0177). C. -Z. Chang acknowledges the support from Gordon and Betty Moore Foundation's EPiQS Initiative (GBMF9063 to C. -Z. C.). C.-Z. Chen acknowledges the support from the National Key R&D Program of China Grant (No. 2022YFA1403700), the Natural Science Foundation of Jiangsu Province Grant (No. BK20230066), and Jiangsu Shuang Chuang Project (JSSCTD202209).

**Figures and figure captions:**

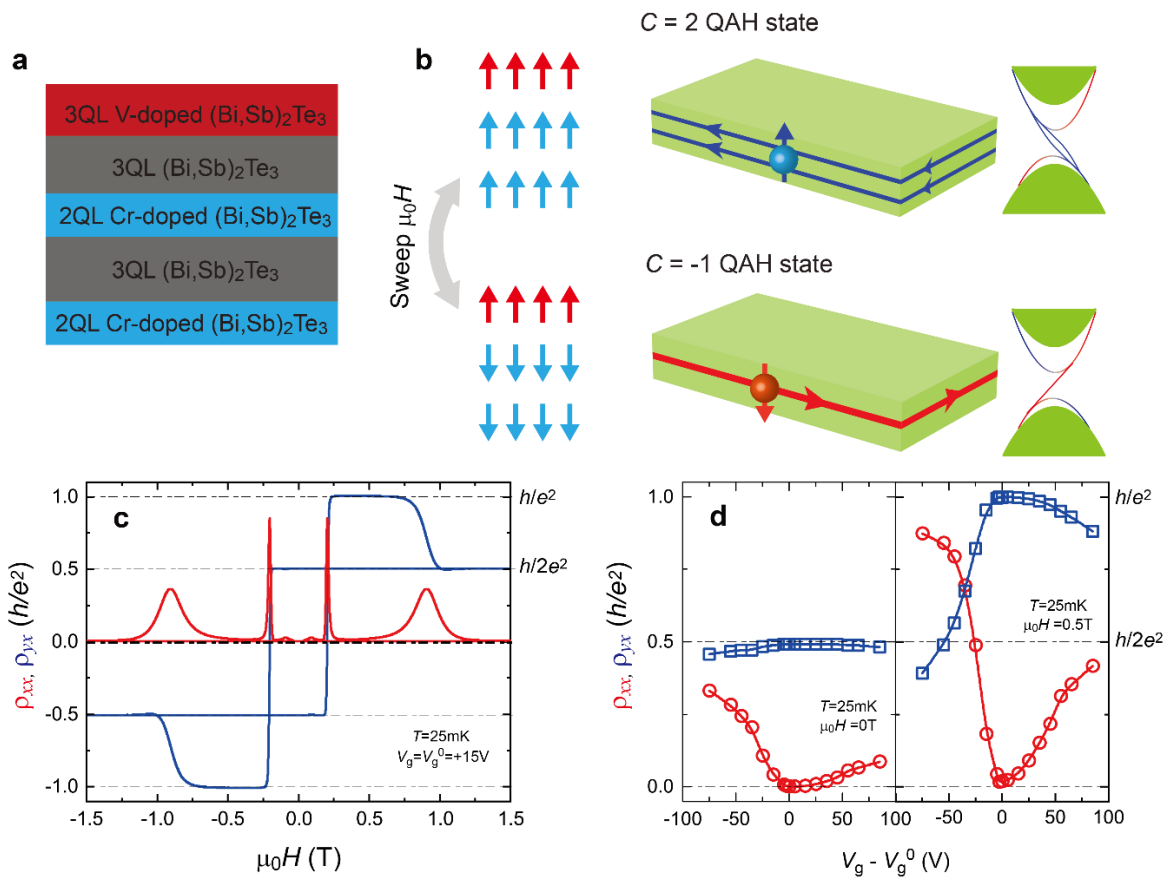

**Fig. 1| MBE-grown 3QL V-doped (Bi,Sb)$_2$Te$_3$/[3QL (Bi,Sb)$_2$Te$_3$/2QL Cr-doped (Bi,Sb)$_2$Te$_3$]$_2$ pentalayer. a,** Side view of the magnetic TI pentalayer. **b,** Schematic of the quantum phase transition between $C=2$ QAH and $C=-1$ QAH states. **c,** $\mu_0H$ dependence of $\rho_{xx}$ (red) and $\rho_{yx}$ (blue) at $V_g=V_g^0=+15V$ and $T=25mK$. **d,** ($V_g$-$V_g^0$) dependence of $\rho_{xx}$ (red) and $\rho_{yx}$ (blue) for the $C=2$ and $C=1$ QAH states at $T=25mK$. The measurements are taken at $\mu_0H=0T$ (left, $C=2$ QAH state) and $\mu_0H=0.5T$ (right, $C=1$ QAH state) after magnetic training at $\mu_0H=1.5T$, respectively.

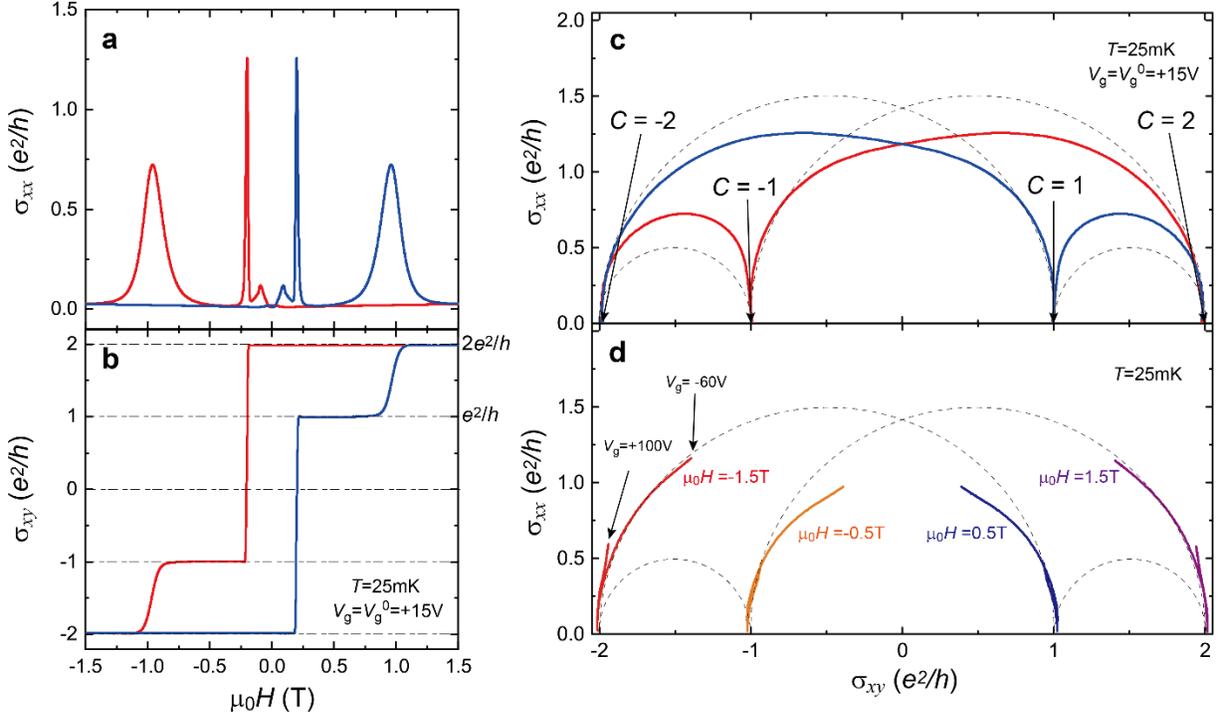

**Fig. 2| Plateau phase transitions with $\Delta C$=1 and $\Delta C$=3 in QAH multilayers. a, b,** $\mu_0 H$ dependence of $\sigma_{xx}$ (**a**) and $\sigma_{xy}$ (**b**) at $V_g=V_g^0$=+15V and $T$=25mK. **c,** Flow diagram ($\sigma_{xy}$, $\sigma_{xx}$) of the pentalayer sample by sweeping $\mu_0 H$. **d,** Flow diagram ($\sigma_{xy}$, $\sigma_{xx}$) of the pentalayer sample by sweeping $V_g$ under different $\mu_0 H$ at $T$=25mK.

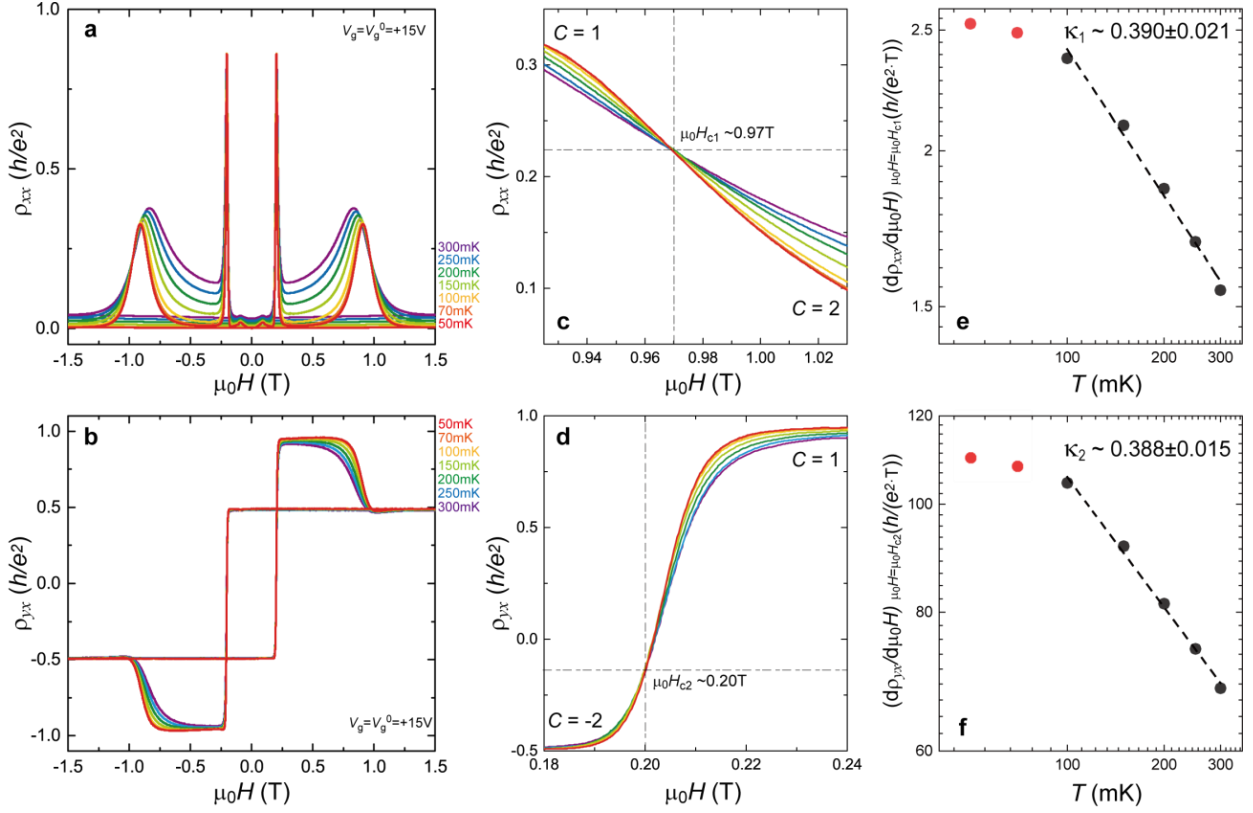

**Fig. 3| Scaling behaviors of plateau phase transitions with Δ*C*=1 and Δ*C*=3 in QAH multilayers. a, b,** $\mu_0 H$ dependence of $\rho_{xx}$ (**a**) and $\rho_{yx}$ (**b**) at different temperatures and $V_g=V_g^0=+15V$. **c,** Enlarged $\mu_0 H$ dependence of $\rho_{xx}$ at different temperatures near $\mu_0 H_{c1} \sim 0.97T$. **d,** Enlarged $\mu_0 H$ dependence of $\rho_{yx}$ at different temperatures near $\mu_0 H_{c2} \sim 0.20T$. **e,** Temperature dependence of $(\frac{d\rho_{xx}}{d\mu_0 H})_{\mu_0 H=\mu_0 H_{c1}}$ on the log scales. **f,** Temperature dependence of $(\frac{d\rho_{yx}}{d\mu_0 H})_{\mu_0 H=\mu_0 H_{c2}}$ on the log scales. The black dashed lines in (**e**) and (**f**) are the linear fits.

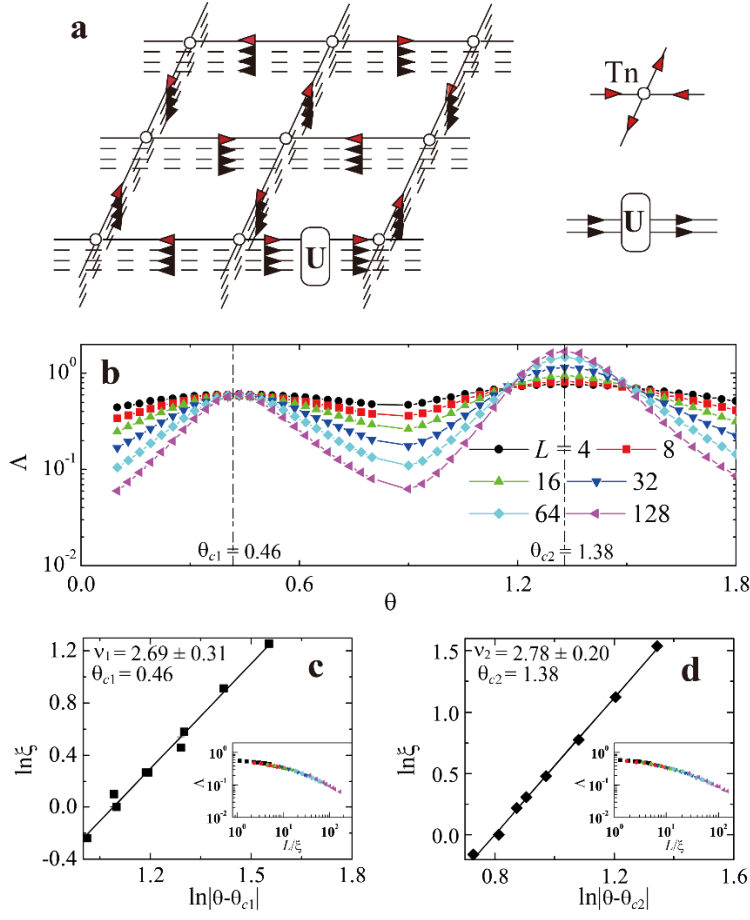

**Fig. 4| Finite-size scaling analysis of the four-layer Chalker-Coddington network model. a,** The four-layer Chalker-Coddington network, with the scattering $T_n$ at each node within the $n^{th}$ layer and the interlayer scattering $U$. **b,** Plot of renormalized localization length $\Lambda$ against $\theta$ for different sizes ($L$=4~128), displaying critical points $\theta_{c1}$=0.46 for the plateau transition with $\Delta C$=1 and $\theta_{c2}$=1.38 for $\Delta C$=3. **c, d,** Logarithmic correlation length $\ln \xi$ against $\ln |\theta - \theta_c|$, fitted with slopes of $v_1$=2.69±0.31 and $v_2$=2.78±0.20. The insets of (**c**) and (**d**) show the data points near the critical point collapse onto a single curve.